\documentclass[conference]{IEEEtran}
\IEEEoverridecommandlockouts
\usepackage{cite}
\usepackage{amsmath,amssymb,amsfonts}
\usepackage{algorithmic}
\usepackage{graphicx}
\usepackage{textcomp}
\usepackage{xcolor}
\def\BibTeX{{\rm B\kern-.05em{\sc i\kern-.025em b}\kern-.08em
    T\kern-.1667em\lower.7ex\hbox{E}\kern-.125emX}}
    
\usepackage{amsmath}
\usepackage{wrapfig}
\usepackage{upgreek}
\usepackage{float}
\usepackage{caption}
\usepackage{subcaption}
\usepackage{tabularray}
\usepackage{pifont}
\usepackage{color}
\usepackage{booktabs}
\definecolor{DustyGray}{rgb}{0.588,0.588,0.588}
\makeatletter
\def\thesection{\arabic{section}}
\def\thesubsection{\thesection.\arabic{subsection}}
\def\thesectiondis{\thesection.}
\def\thesubsectiondis{\thesection.\arabic{subsection}.}
\makeatother

\begin{document}
\title{SuperSNN: A Hardware-Aware Framework for Physically Realizable, High-Performance Superconducting Spiking Neural Network Chips
\thanks{This work is supported by a grant from the National Science Foundation (NSF) under the project Expedition: Discover (Design and Integration of Superconducting Computation for Ventures beyond Exascale Realization), grant number 2124453, and by the Department of Energy (DOE) AI4Science program with project name Flexbrain.}}

\author{\IEEEauthorblockN{Changxu Song, Arda Caliskan, Beyza Zeynep Ucpinar, Yasemin Kopur, Mustafa Altay Karamuftuoglu, \\ Sasan Razmkhah, Shahin Nazarian, Massoud Pedram}
\IEEEauthorblockA{University of Southern California, Los Angeles, USA \\
\{changxus, acaliska, ucpinar, karamuft, razmkhah, shahin.nazarian, pedram\}@usc.edu}
}

\maketitle

\begin{abstract}
Despite numerous proposed designs for superconducting neural networks (SNNs), most have overlooked practical fabrication constraints, leading to implementations limited to only a few neurons or synapses. Current superconducting technologies, such as MIT LL SFQ5ee, impose severe limitations on chip area, routing, and input/output pin counts (e.g., 5x5 $mm^2$ chip with 40 pins), drastically restricting network size and complexity. These hardware constraints necessitate a comprehensive framework to tailor network designs for physical realizability while minimizing accuracy loss.
This paper introduces SuperSNN, a comprehensive framework for the implementation of full superconducting SNNs on a chip within these constraints. The key technical contributions include: (1) A hardware-aware training methodology for SNNs, utilizing off-chip pruning and weight quantization for energy-efficient superconducting implementations. (2) Design and layout of an inference SNN chip that incorporates novel high fan-in neurons and custom superconducting cells. (3) An optimized locally synchronous, globally synchronous (LAGS) clock distribution scheme for robust circuit implementation and management of data transfer delays in SFQ SNNs.
The main results and findings demonstrate the effectiveness of the framework: (1) The complete network achieved 96.47\% accuracy on the full MNIST dataset after quantization and pruning. (2) The fabricated SuperSNN chip successfully classified a reduced set of digits (2, 3, and 4) with 80.07\% accuracy, reaching a maximum of 86.2\% accuracy for digits 0, 1, and 2. (3) The chip operates at an ultra-high 3.02 GHz clock frequency. (4) It occupies a compact area of 3.4 × 3.9 $mm^2$, incorporates 5,822 Josephson Junctions, consumes 2.15 mW static power, and has an exceptionally low energy cost of 6.55 fJ (or $1.31 \times 10^{-6}$ nJ) per inference.

\end{abstract}


\section{Introduction}
Neuromorphic computing seeks to fundamentally transform data processing by mimicking the operation of biological nervous systems. Traditional computers rely on the Von Neumann architecture, which suffers from significant inefficiencies due to the physical separation of memory and processing units, necessitating continuous data transfer. Although alternative neural network architectures have been proposed to enhance performance, conventional approaches such as Artificial Neural Networks (ANNs), which depend heavily on computationally intensive convolution and matrix multiplication, face exponential increases in computational demands and power consumption as network size grows.

To bridge this energy gap, brain-inspired Spiking Neural Networks (SNNs) have been proposed \cite{Hardware_implementation_of_spiking_neural_networks_on_FPGA}. SNN emulates the efficiency of the human brain, offering superior energy performance compared to ANN. This efficiency comes from their event-driven processing paradigm and binary spike activations. Unlike ANNs that continuously process data streams, SNNs are triggered only by input events and remain inactive otherwise; their binary spikes require only simple integer accumulations instead of multiply-accumulate operations, reducing latency and power consumption. In addition, low-precision temporal coding and the inherent sparsity of spike trains further minimize memory traffic, enhancing both speed and energy efficiency. When implemented on specialized neuromorphic hardware, this event-driven sparsity significantly reduces data movement between on-chip and off-chip resources.

Despite the many benefits of SNNs, exploring suitable alternative hardware architectures for them remains a challenge. Although neuromorphic optical computing offers low power and latency \cite{Optical_neural_networks_progress_and_challenges}, reconfigurability and integration challenges make complex optical systems less practical compared to CMOS-based solutions. Memristor-based devices provide a CMOS-compatible option and are well suited for synaptic emulation due to their inherent variability \cite{Power-efficient_combinatorial_optimization_using_intrinsic_noise_in_memristor_Hopfield_neural_networks}. However, they are analog devices requiring high precision, making noise unavoidable, and their dissipative nature during read/write operations presents a serious heating problem as systems scale.

Superconducting neuromorphic architectures offer a scalable solution to these issues, providing significantly improved energy efficiency compared to conventional electronic systems. Operating at cryogenic temperatures, they enable ultrahigh-speed signal propagation with near-zero resistive losses, ensuring zero data degradation \cite{Beyond-CMOS}. Digital Single-Flux-Quantum (SFQ) circuits, leveraging superconducting technology, are asynchronous and spike-based, operating at tens of GHz with ultra-fast switching characteristics. They transmit data as quantized pulses at high speeds. These characteristics make SFQ circuits a promising candidate for emulating the event-based nature and energy efficiency of SNN processing.

A broad spectrum of superconducting approaches has been explored to realize the advantages of SNNs using SFQ or similar superconductor logic families. Early work demonstrated mimicking spiking neuron behavior with Josephson Junctions (JJs), including leaky integrate-and-fire models simulated via SFQ pulse generation and thresholding \cite{Hirose2006}. Further investigations showed synchronization and coupling dynamics between JJ-based neurons on the picosecond time scale \cite{Segall2017}. Developments focused on biologically plausible dynamics, with circuits fabricated that emulate action potentials, refractory periods, and thresholds, demonstrating high-speed and energy-efficient behavior \cite{Segall2010}. The hybrid integration of photonics and superconductors emphasized the scaling potential \cite{Shainline2024}. Magnetic JJs (MJJs) facilitated tunable and stochastic synapses with sub-attojoule operation and analog weight updates \cite{Russek2016, Schneider2018}. Various neuron implementations, such as nanowire-based designs \cite{Toomey2019} and quantum phase-slip junction (QPSJ) designs \cite{Cheng2018, Cheng2021}, offered low-power switching and high fan-in. The hybrid RSFQ-QFP circuits introduced dynamic weight programming \cite{Jardine2023}. Architecture-level advancements included multilayer spike rate coded networks using purely SFQ logic, preserving asynchronous, ultralow-power operation \cite{Edwards2024}. The bioSFQ framework bridged the analog and digital domains, enabling SFQ-based analog operations \cite{Semenov2023, Golden2025}. The learning mechanisms were incorporated through local STDP rules \cite{Segall2023, Segall2025}, trainable neurons \cite{Ucpinar20}, and unsupervised networks \cite{Karamuftuoglu2024}. Integration efficiency techniques such as ternary weights and temporal encoding reduce the area and the count of components \cite{Han2024, Li2024}. Processors such as SUSHI \cite{Liu2023} and an ultra-high-speed processor demonstrated RSFQ / SFQ-based platforms with on-chip learning support \cite{Bozbey2020}.

Superconductor logic still faces significant challenges in physical circuit implementation. 
Although in-memory computing architectures such as SNNs partially address the lack of dense memory, a key challenge remains: implementing neurons with high fan-in and fan-out. Splitter circuits handle fan-out in SFQ. However, high fan-in is more complex. \cite{Karamuft_Scalable} introduced a superconductor neuron design using mutual coupling branches for the dynamics of somatic operation, which was used to design a SNN feedforward inference. The proposed network achieved 95.13\% accuracy after pruning and quantization.

Despite this progress, a fully fabricated and inference-ready feedforward SNN implementation has not yet been realized. Given the potential for tens of GHz classification speeds, ultra-low power consumption, and a minimal footprint, developing such a chip remains a valuable goal. Although software implementations can achieve high accuracy on small datasets such as MNIST, the fabrication process imposes significant constraints. Current technologies, such as the MIT LL SFQ5ee process \cite{Tolpygo2016_MITLL}, feature a limited number of routing metal layers, a restricted chip area of $5\times 5 mm^2$, and a small number of input/output pads (40 pads placed uniformly around the periphery of the chip), severely limiting the size of the networks that hardware can support. Furthermore, data transfer delays require careful synchronization design. These hardware limitations mean that the network must be tailored to physical constraints, often through compaction techniques with minimal loss of accuracy. Consequently, designing an SNN chip based on SFQ logic within these constraints presents substantial challenges.

In this work, we present a comprehensive framework for developing on-chip inference SNNs that comply with hardware constraints of superconductor electronics and successfully realize the chip layout. We utilize our newly designed SFQ standard cell libraries for manual layout, routing, and optimization. We introduce SuperSNN, an end-to-end framework for implementing fully superconducting SNNs on a chip using the MIT LL SFQ5ee fabrication process. Our framework encompasses network training, pruning to limit fan-in and fan-out, and weight quantization to ensure physical realizability. To demonstrate the effectiveness of our approach, we design, train, and implement an SNN for MNIST classification on a single 5$\times$5 mm$^2$ chip. We employ a novel high fan-in neuron design optimized for scalability and utilize hybrid SFQ wiring cells for synaptic connections. The proposed architecture leverages the event-driven nature of SNNs and the ultra-high-speed, low-power advantages of SFQ circuits to overcome the inefficiencies present in conventional systems.

The main contributions of this paper are:
\begin{itemize}
\item Developing a hardware-aware training methodology for SNNs optimized for energy-efficient superconducting implementations. This framework enables off-chip training with pruning and weight quantization, which is crucial for physical realizability.
\item Design of an inference SNN chip that incorporates high-fan-in neurons and custom superconducting cells. This resulting network is designed to operate at a clock frequency of 3.02 GHz and occupies a compact area of 3.4$\times$3.9 $mm^2$.
\item Implementation of a locally asynchronous, globally synchronous (LAGS) clock distribution scheme to ensure an optimized and robust circuit implementation of the SFQ SNN. This scheme helps address the challenge of data transfer delays in superconducting logic.
\end{itemize}

\section{Background}

\subsection{Leaky Integrate-and-Fire Neuron Model}
Neurons are surrounded by a lipid membrane that electrically isolates the cell while allowing ion flow through gated channels. This membrane behaves like a capacitor, storing and integrating charge. The leaky integrate-and-fire (LIF) model captures this dynamic by abstracting neuronal behavior as temporal integration of inputs, rather than applying a direct activation function. The LIF neuron models membrane integration as an RC-like circuit, accumulating inputs until the membrane potential exceeds a threshold and triggers a spike. Rather than modeling the spike waveform, the LIF model treats it as a discrete event, striking a balance between biological realism and computational simplicity. We adopt a simplified LIF model from \texttt{snnTorch} with a hard reset mechanism. This reset period effectively serves as a refractory phase, allowing for temporal filtering and contributing to spike-based computation.
The spiking output is defined by the Heaviside function in Eq.~\ref{eq:spike}:

\begin{equation}
S[t] =
\begin{cases}
1, & \text{if } U[t] > U_{\text{thr}} \\
0, & \text{otherwise}
\end{cases}
\label{eq:spike}
\end{equation}
where $S[t]$ is the binary spike output at time $t$, $U[t]$ is the membrane potential, and $U_{\text{thr}}$ is the threshold. Each time step $t$ becomes discrete, and at most one spike may occur.
The membrane potential update rule is given in Eq.~\ref{eq:membrane}:
\begin{equation}
U[t+1] = \beta U[t] + WX[t+1] - S[t]U_{\text{thr}}
\label{eq:membrane}
\end{equation}
Here, $\beta$ is a decay constant ($0 < \beta < 1$) that models the exponential leakage of the potential in the absence of input. $W$ is the synaptic weight and $X[t+1]$ is the input spike in the next time step. The subtraction term $S[t]U_{\text{thr}}$ resets the membrane after a spike, enforcing the hard reset behavior. This formulation compactly models both integration and decay, closely resembling the dynamics of an RC circuit with time constant $\tau = RC$.

\subsection{Training Methodology of The Network}
\begin{figure}
    \centering
    \includegraphics[width=\columnwidth]{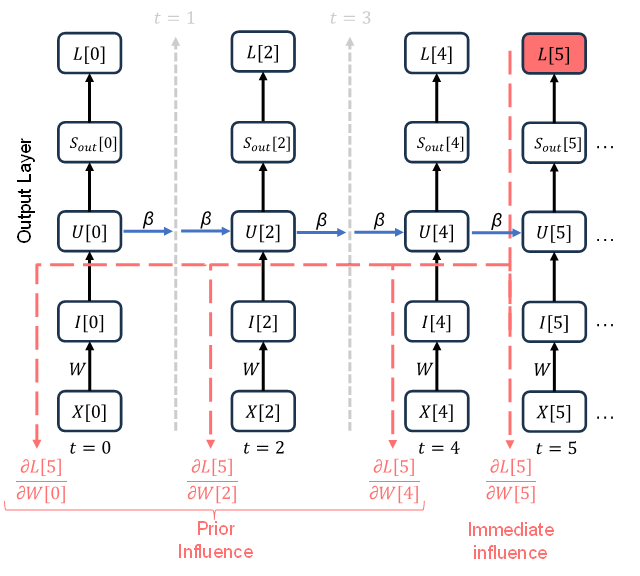}
    \caption{Unrolling of a spiking neural network over multiple time steps.}
    \label{fig:FigUnrollingSNN}
\end{figure}

When training neural networks composed of discrete, recurrent spiking neurons, these networks may be approximated as Recurrent Neural Networks (RNNs). Specifically, the computational graph of the spiking neurons over multiple time steps is first unrolled, as shown in Figure~\ref{fig:FigUnrollingSNN}, followed by the application of backpropagation through time (BPTT).

At each time step, the loss gradient is computed with respect to the shared weights, and these gradients are summed across all time steps to obtain the total gradient for parameter updates. The gradient with respect to the weight parameter $W$ can be expressed as:
\begin{equation}
\frac{\partial \mathcal{L}}{\partial W} 
= \sum_{t} \sum_{s \leq t} \frac{\partial \mathcal{L}[t]}{\partial W[s]},
\label{eq:gradient_sum}
\end{equation}
Here, $\mathcal{L}[t]$ denotes the loss in time step $t$, and $W[s]$ refers to the shared synaptic weight in time $s$. During backpropagation over time, a key step is computing the loss gradient at time $t$ with respect to the weight at the previous step, $W[t{-}1]$. Assuming that the loss depends directly on spikes and is differentiable, this gradient is given by Eq.~\ref{eq:grad_w_t_minus_one}:

\begin{equation}
    \begin{gathered}
        \frac{\partial \mathcal{L}[t]}{\partial W[t-1]} 
        = \frac{\partial \mathcal{L}[t]}{\partial S[t]}
              \cdot \frac{\partial \tilde{S}[t]}{\partial U[t]} \\
        \cdot \frac{\partial {U}[t]}{\partial U[t-1]}
              \cdot \frac{\partial U[t-1]}{\partial I[t-1]}
              \cdot \frac{\partial I[t-1]}{\partial W[t-1]}
    \end{gathered}
    \label{eq:grad_w_t_minus_one}
\end{equation}
where $\tilde{S}[t]$ is a surrogate for the nondifferentiable spike output $S[t]$, allowing gradient-based optimization. $U[t]$ is the membrane potential and $I[t{-}1]$ is the input current at time $t{-}1$, depending on the weight $W[t{-}1]$. These quantities are linked via the LIF update rule in Eq.~\ref{eq:membrane}.

The partial derivatives in Eq.~\ref{eq:grad_w_t_minus_one} arise directly from the membrane update rule and basic calculus. Their values are as follows:
\begin{itemize}
    \item \( \partial U[t] / \partial U[t{-}1] = \beta \), the decay factor, representing the influence of the previous membrane potential.
    \item \( \partial U[t{-}1] / \partial I[t{-}1] = 1 \), from the direct relationship between the input current and the membrane potential.
    \item \( \partial I[t{-}1] / \partial W[t{-}1] = X[t{-}1] \), where \( X[t{-}1] \) is the pre-synaptic input spike at time \( t{-}1 \).
\end{itemize}

Applying the chain rule, the immediate effect of the weight \( W[t{-}1] \) on the membrane potential \( U[t] \) is \( \beta \cdot X[t{-}1] \).
When spikes are used for training, \( S[t] \) is typically defined by the Heaviside function, which is nondifferentiable at \( x = 0 \) and has zero derivative elsewhere, resulting in undefined or vanishing gradients. To avoid this, frameworks like \texttt{snnTorch} apply surrogate gradients, replacing the Heaviside function with an approximation during backpropagation. For instance, the arctangent surrogate yields \( \partial \tilde{S} / \partial U = 1 / \left[\pi \left(1 + (U\pi)^2\right)\right] \), allowing stable, nonzero gradients.

Alternatively, if the loss is defined over the membrane potential instead of spikes, gradients can bypass \( S[t] \) entirely, as shown in Eq.~\ref{eq:grad_w_t_minus_one_membrane_potential}:
\begin{equation}
    \frac{\partial \mathcal{L}[t]}{\partial W[t-1]}
    = 
    \frac{\partial \mathcal{L}[t]}{\partial U[t]}
    \cdot
    \frac{\partial {U}[t]}{\partial U[t-1]}
    \cdot
    \frac{\partial U[t-1]}{\partial I[t-1]}
    \cdot
    \frac{\partial I[t-1]}{\partial W[t-1]}
    \label{eq:grad_w_t_minus_one_membrane_potential}
\end{equation}

In supervised multiclass classification, the predicted label corresponds to the output neuron with the highest spike count. To guide the network toward this behavior during training, we encourage the correct class neuron to spike more frequently by increasing its membrane potential beyond the threshold (\( U > U_{\text{thr}} \)), while suppressing the potentials of incorrect class neurons (\( U < U_{\text{thr}} \)).
This mechanism is implemented by applying a cross-entropy loss to the membrane potentials of the output layer, referred to as \emph{mem loss}. As shown in Eq.~\ref{eq:softmax_membrane}, we first apply a softmax function to the membrane potentials:
\begin{equation}
    p_i[t] = \frac{e^{U_i[t]}}{\sum_{j=0}^C e^{U_j[t]}}
    \label{eq:softmax_membrane}
\end{equation}
where \( C \) denotes the number of classes and \( U_i[t] \) is the membrane potential of the \( i \)th output neuron at time step \( t \). We compute the softmax probabilities \( p_i[t] \) over membrane potentials, and define the target label \( y_i \) as a one-hot encoded vector (Eq.~\ref{eq:target_delta}). The cross-entropy loss is then computed between \( p_i[t] \) and \( y_i \), as shown in Eq.~\ref{eq:crossentropy_t}:
\begin{equation}
    y_i = 
    \begin{cases}
    1, & i = \text{target class},\\
    0, & i \neq \text{target class}
    \end{cases}
    \label{eq:target_delta}
\end{equation}

\begin{equation}
    \mathcal{L}_{CE}[t] = -\sum_{i=0}^{C} y_i\log(p_i[t])
    \label{eq:crossentropy_t}
\end{equation}

This loss is applied at each time step, resulting in a total loss that is the sum over all time steps, as shown in Eq.~\ref{eq:total_crossentropy}:
\begin{equation}
    \mathcal{L}_{CE} = \sum_{t} \mathcal{L}_{CE}[t]
    \label{eq:total_crossentropy}
\end{equation}
The \emph{mem loss} encourages a higher membrane potential for the correct class and suppresses it for incorrect classes, thus promoting spiking activity in the target neuron while inhibiting others. This loss is typically sufficient; however, under extreme constraints, such as reducing the network to a single forward pass and fewer than 25 neurons for deployment on specialized SNN hardware, \emph{mem loss} alone may cause under-fitting.

To address this, we introduce an additional \emph{spike loss}, forming a hybrid objective aligned with Multi-task Learning \cite{Multi-task_Learning_Using_Uncertainty_to_Weigh_Losses_for_Scene_Geometry_and_Semantics,sener2019multitasklearningmultiobjectiveoptimization}. Like \emph{mem loss}, it applies cross-entropy, but to spike outputs \( S[t] \). We compute a softmax over spike counts in the output layer and use the same one-hot target comparison, thereby encouraging the correct neuron to spike more often while suppressing spiking in others.
The formulation mirrors that of \emph{mem loss}, with the only difference being the substitution of membrane potentials \( U_i[t] \) with spike outputs \( S_i[t] \) in Eq.~\ref{eq:softmax_membrane}, resulting in a parallel cross-entropy loss computed over binary spiking activity.

\subsection{Circuit Components}

\subsubsection{Transmission Lines}
A Josephson Transmission Line (JTL) consists of a series of grounded, shunted, and biased JJs that are connected by inductors, which are configurable for different design requirements, such as transmission latency or peak SFQ pulse amplitude \cite{Josephson-Junction_Technology_for_Sub-Terahertz-Clock-Frequency_Digital_Systems}. 
To simplify both the design and analysis, the bias currents and inductance are chosen to be identical in this design. 

In addition to providing a significant margin, JTLs have drawbacks, such as occupying a large area and having limited wiring distance. Although the switching power is minimal for a JTL, the cumulative power consumption becomes considerable for long-distance connections operating at frequencies greater than 10~GHz. Consequently, Passive Transmission Lines (PTL) that allow SFQ pulses to propagate at speeds approaching that of light with minimal signal attenuation are often used instead for designs that involve long distances. A PTL typically consists of a stripline connecting a driver and a receiver. 

\subsubsection{D Flip-Flop (DFF)}
A destructive readout D flip-flop (DFF) typically consists of three Josephson junctions. To store a single flux quantum in the loop, the inductance $L$ must satisfy the condition $L > \Phi_{0}/I_{\text{C}}$, which can be maintained indefinitely. A clock pulse is applied at the input terminal to read the state of the flip-flop, to redirect the stored current to the output, and generate an SFQ pulse.

\subsubsection{Input Shift Register}
\begin{figure}
    \centering
    \includegraphics[width=0.9\linewidth]{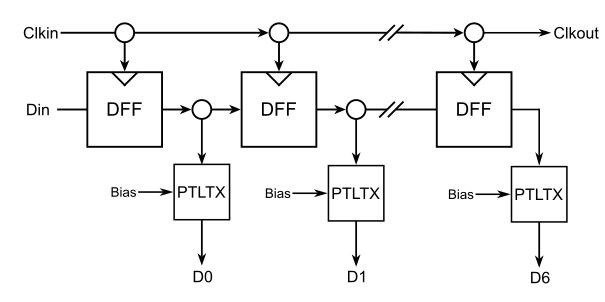}
    \caption{Input shift register circuit. The clock follows the data path and propagates the data through the DFFs.}
    \label{fig:enter-label}
\end{figure}

To meet the 40-pin constraint of the chip, we implemented a shift register (SR) input scheme to minimize pin usage. The input layer consists of seven shift registers, each with seven DFFs in series, resulting in 49 total DFFs. The input SR layer allows a downsampled \(7\times 7 \) image to be loaded bit-by-bit over seven data pins in seven clock cycles. Each register shares a common serial clock, implemented by chaining local clocks, enabling all registers to be driven from a single clock pin. As shown in Fig.~\ref{fig:enter-label}, each DFF output is split—one path connects to the next DFF, and the other feeds the next layer. Biasing for the next layer is disabled until all bits are loaded, preventing premature activations.

The clock tree is implemented serially to reduce area overhead. This design introduces a 433 ps delay across the entire shift register path, limiting the system's throughput to just over 1 GHz. By contrast, an H-tree-based clock structure would shift the bottleneck to neurons and increase maximum throughput to 8.9~GHz.

\subsubsection{Fan-out Circuit}
SNNs can require large fan-outs while propagating data throughout the network. We created fan-out circuits built using cascaded splitters (SPL) cells of fan-out two to facilitate this requirement. The maximum fan-out is seen to be nine, which corresponds to eight cascaded SPL cells. In our library, each SPL cell introduces 3.1 ps of delay, resulting in approximately 24.8 ps of delay from the fan-out circuit.

\subsubsection{High Fan-in Neural Cell}
\begin{figure}
    \centering
    \includegraphics[width=0.9\linewidth]{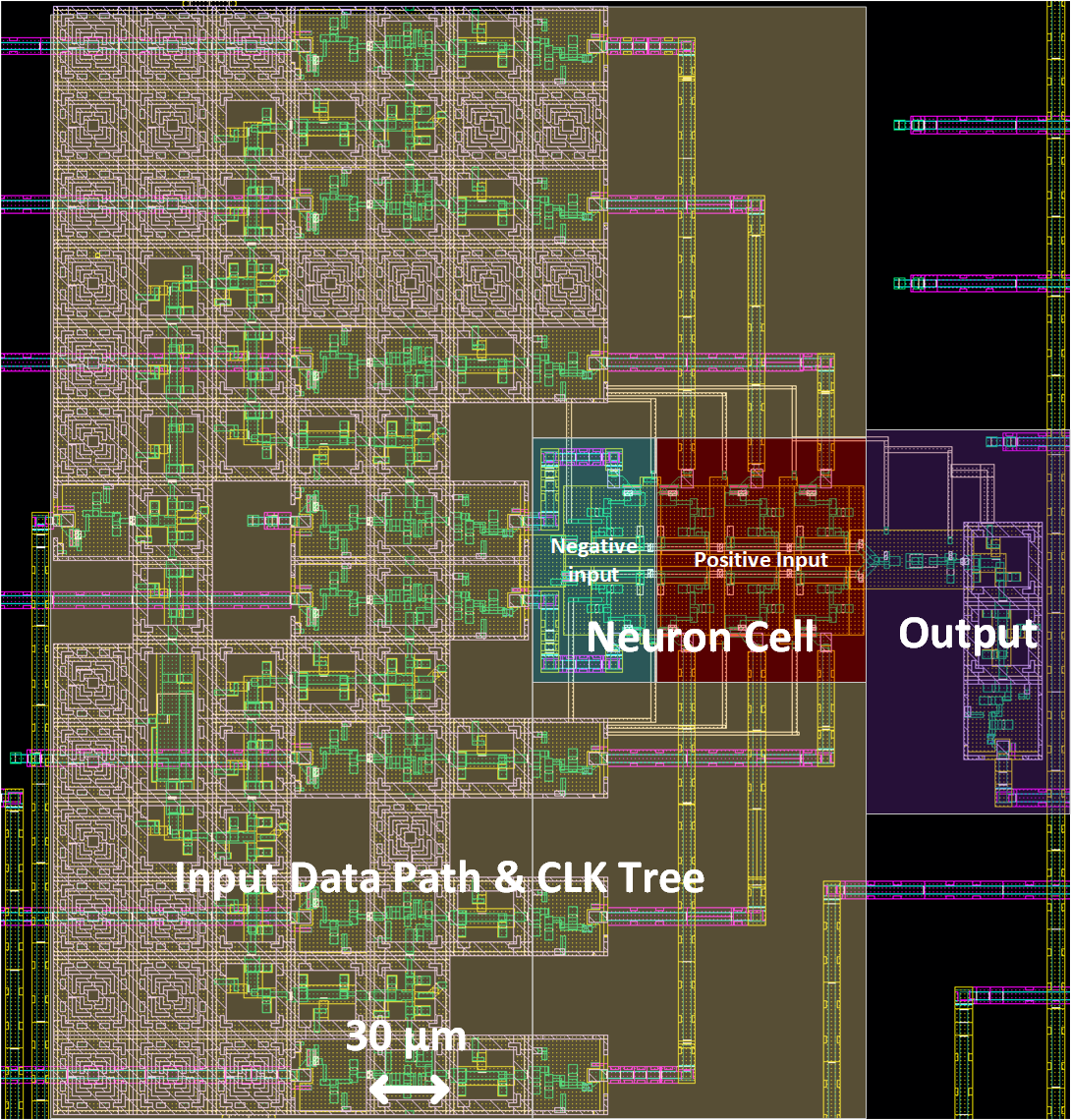}
    \caption{NeuronLayout}
    \label{fig:FigNeuronLayout}
\end{figure}

Each neuron receives input current via mutual inductive coupling through multiple dendrites, each weighted as \( +1 \), \( -1 \), or 0. When the combined loop current and bias exceed the JJ's critical current, the neuron fires an SFQ pulse, which is transmitted to downstream neurons. For output neurons, the SFQ signal is converted to a DC level and routed to an output pin. For data synchronization, each input path includes a DFF, avoiding the need for manual delay balancing. A JTL is inserted on the positive path to introduce latency, ensuring that the negative pulse arrives first \cite{Karamuft_Scalable}. Fig.~\ref{fig:FigNeuronLayout} shows the layout of an eight-input neuron with two dendritic branches, six positive and two negative. Synchronized input pulses propagate through the coupling network before reaching the neuron.

In the original neuron peripheral design, JTLs were used to interface with both fan-in and fan-out nodes before connecting to PTLs. In our implementation, the PTL receivers are directly integrated at the fan-in nodes, allowing the neuron and PTL interface to be optimized together as a single unit. To improve parametric yield, a wider timing margin is allocated for incoming pulses, enabling greater tolerance to variations in signal arrival times. However, this increased timing margin requires a lower decay rate in the leaky loop of the neuron, which ultimately reduces the maximum operating frequency.
\begin{figure*}[!t]
\centering
\begin{subfigure}{0.65\linewidth}
    \centering
    \includegraphics[width=\linewidth]{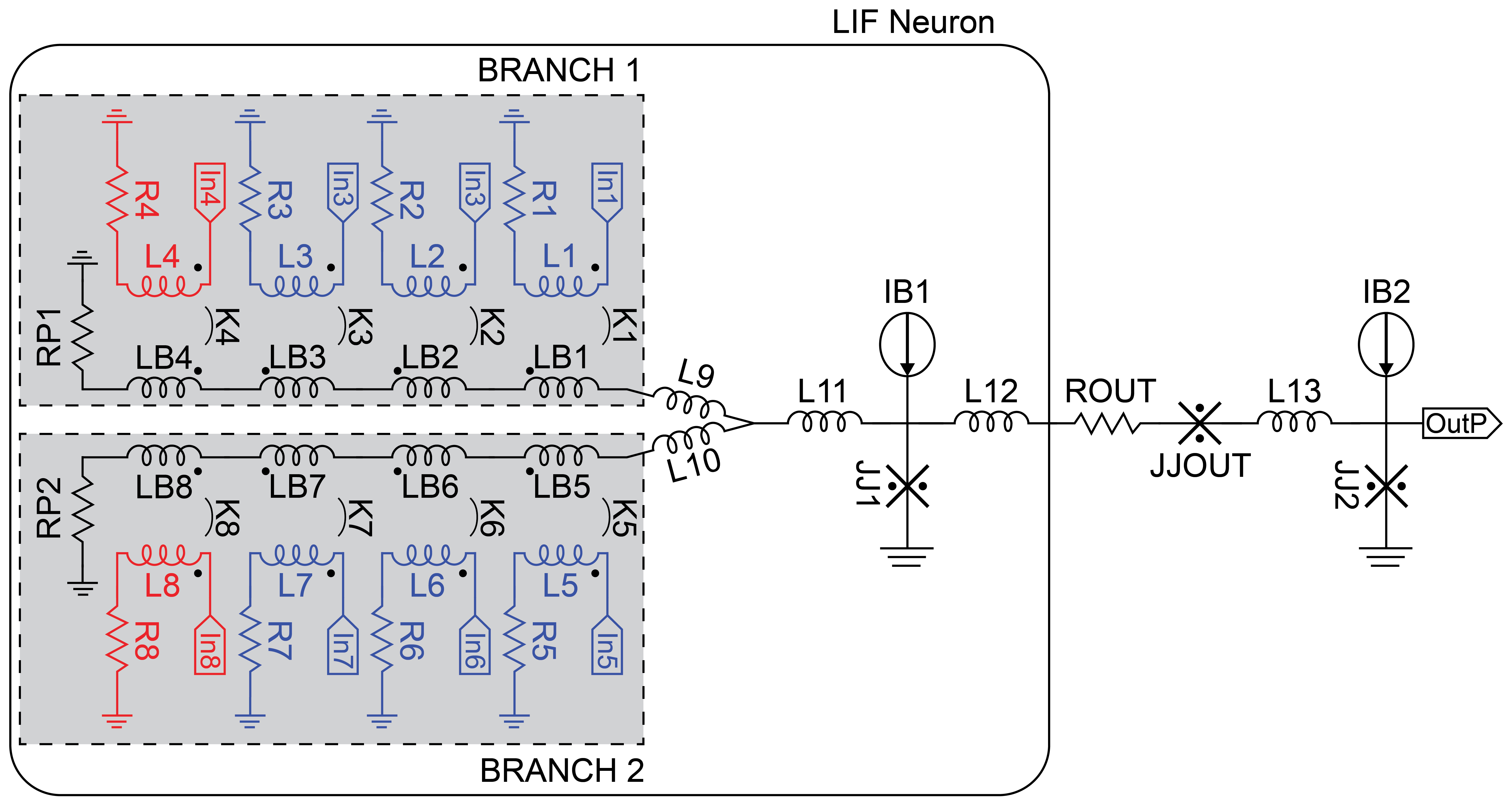}
    \caption{Schematic of high fanin neuron \small(L1 = L2 = L3 = L4 = L5 = L6 = L7 = L8 = 11.52 pH, LB1 = LB2 = LB3 = LB4 = LB5 = LB6 = LB7 = LB8 = 11.52 pH, R1 = R2 = R3 = R5 = R6 = R7 = 0.56 $\Omega$, R4 = R8 = 0.40 $\Omega$, K1 = K2 = K3 = K5 = K6 = K7 = 0.6, K4 = K8 = -0.6, L9 = L10 = 1.10 pH, RP1 = RP2 = 0.55 $\Omega$, L11 = 0.93 pH, $I_{JJ1c}$ = 0.1 mA, L12 = 0.5 pH, ROUT = 4.32 $\Omega$, $I_{JJOUTc}$ = 0.15 mA, L13 = 1.90 pH, $I_{JJ2c}$ = 0.17 mA)}
    \label{fig:neuronSimSch}
\end{subfigure}
\hfill
\begin{subfigure}{0.30\linewidth}
    \centering
    \includegraphics[width=\linewidth]{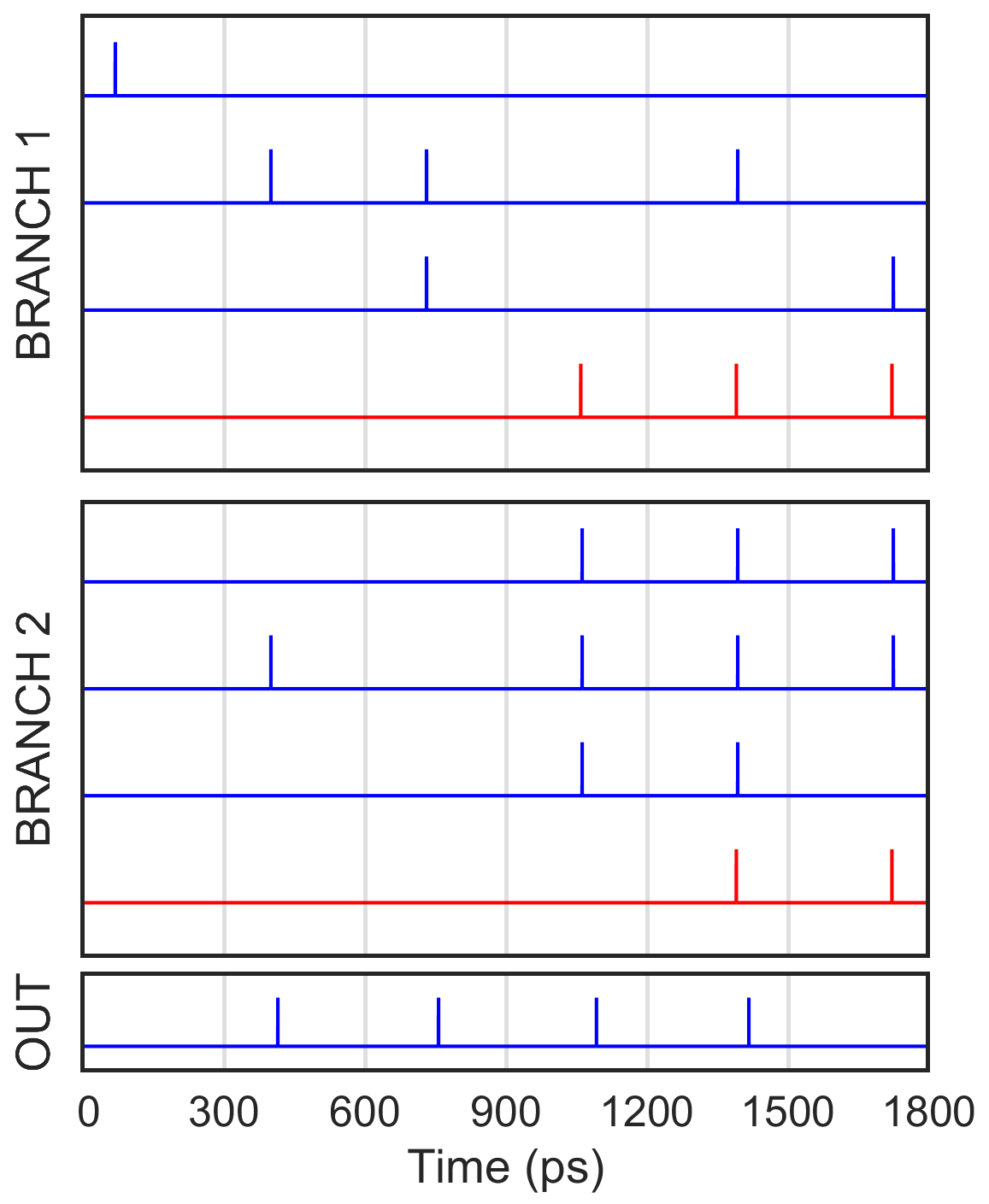}
    \caption{Simulation result of the network showing maximum operating at a clk frequency of 3.02 GHz with peripheral JJ pulse observations.}
    \label{fig:neuronSimResult}
\end{subfigure}
\caption{High fan-in neuron design. Each neuron receives weighted inputs via two dendritic bundles—BRANCH 1 (RP1, synapses L1-L4) and BRANCH 2 (RP2, synapses L5-L8)—and fires an SFQ pulse once its combined current surpasses the JJ's critical current; in the schematic, excitatory (positive-weight) inputs are shown in blue, while inhibitory (negative-weight) inputs are shown in red.}
\label{fig:FigHifaninneuron}
\end{figure*}


\subsubsection{Network Architecture}
Using the described circuit components, we constructed a feedforward SNN composed of SPL cells, PTLs, and neuron cells. The input layer consists of 49 DFFs, followed by 18 and three neurons in the hidden and output layers, respectively. Inputs are loaded through a shift register and routed through SPL trees and PTLs, expanding 49 inputs into 97 outputs. These are received by 18 neuron cells, which use internal clock trees and DFFs for synchronization before coupling to the soma to trigger activation. Each neuron outputs a single bit, transmitted via PTL links to a second fan-out block that converts 18 inputs into 23 outputs. These are routed to the final three output neurons, whose spikes are converted to DC and routed off-chip to represent the classified digit.

\section{Software Framework}

\subsection{Applying the Training Methodology on the SNN Network}
The proposed training methodology demonstrates exceptional versatility and effectiveness, achieving high accuracy across both conventional and highly constrained superconductor electronics hardware and tapeout capabilities. In this study, we validate the approach using two variants of the inference network. The first, known as the complete network, performs classification on digits 0–9. Each neural cell supports up to 64 inputs in this configuration, without requiring uniformity across cells. There are no constraints on space or pin count, and output spikes from each neuron in the final layer are counted to determine the classification result. The second, called the chip network, is designed to operate under real superconducting hardware constraints. These include a chip with only 40 pins (7 input and three output pins) and limited space accommodating just 25 neurons. As a result, classification is restricted to digits 2, 3, and 4. In addition, the chip cannot count the total number of output spikes. Therefore, the chip network must simulate the behavior of an inference SNN within all these hardware constraints.


As we will demonstrate, both networks achieve excellent accuracy despite their respective constraints, validating the robustness of our training approach.

\begin{figure}
    \centering
    \includegraphics[width=0.9\linewidth]{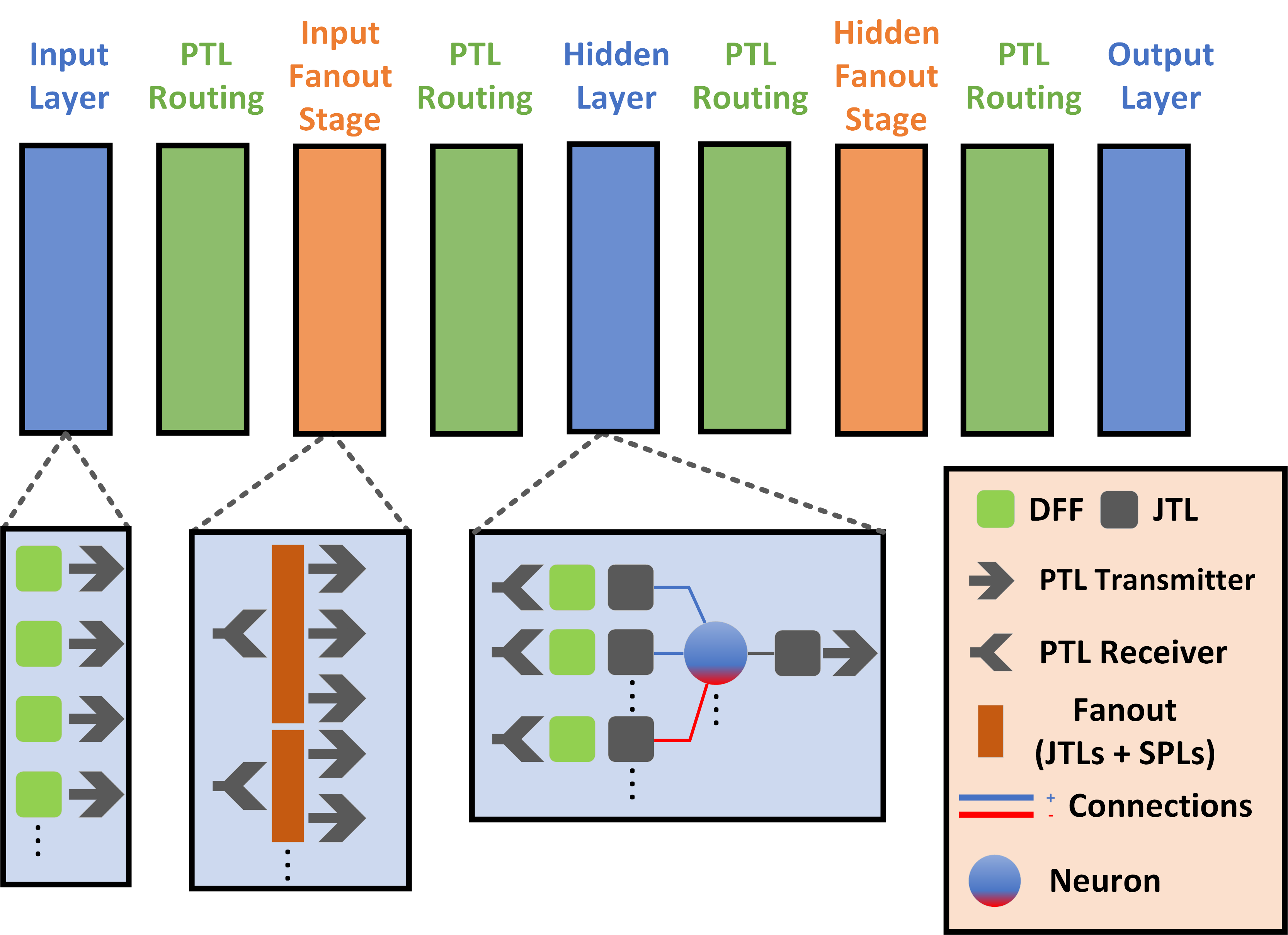}
    \caption{
        Network Architecture. Inputs spread through a fan-out stage into the hidden layer; their single-bit outputs fan-out again using PTL/JTL links, finally converging on the output layer, with DFFs coordinating timing throughout.
        }
    \label{fig:FigNetwork}
\end{figure}

\subsubsection{Training the Complete Network}
The complete network consists of five layers: an input layer, three hidden layers, and an output layer. The input layer comprises $28 \times 28$ registers, while the hidden layers contain 128, 96, and 96 neurons, respectively. The output layer consists of 10 neurons. For data processing, each $28 \times 28$ pixel training sample is flattened and presented to the network repeatedly at each time step (with a simulation duration of 25 time steps). During inference, input data is quantized to 1 bit to emulate the presence or absence of an SFQ pulse. The weights are limited to values of $+1$, $-1$, and $0$ by enforcing discrete coupling coefficients $K_i\in\{+1,-1,0\}$ between each primary inductor $L_i$ and its magnetically coupled secondary $LB_i$ As shown in  Fig.~\ref{fig:FigHifaninneuron}. , so that the induced current in every active secondary coil $LB_i$, which flows into the JJ, has identical magnitude, differing only in sign. Each neuron's fan-in is limited to 64, without specific constraints on the distribution between positive- and negative-weight connections. 
For output classification, we employ spike count decoding \cite{Unsupervised_learning_of_digit_recognition_using_spike-timing-dependent_plasticity,Unsupervised_SFQ-Based_Spiking_Neural_Network}, where network prediction is determined by counting the spikes emitted by each output neuron throughout the simulation window. The neuron with the highest spike count indicates the predicted class.

The entire training process for this network is divided into five stages. Each stage runs for a predetermined number of epochs, with the model yielding the highest accuracy on the full test dataset being preserved.

\paragraph{Unrestricted training} 
In this stage, the network is allowed to train freely using a membrane potential-based loss (mem loss) with the AdamW optimizer on 32-bit floating point data. The pre-binarized \(28\times28\) input vector is fed directly into the network, and each vector is propagated over 25 time steps, yielding 25 prediction results whose output-layer membrane potentials are aggregated to compute the overall mem loss, which is then backpropagated to update the weights. 

\paragraph{Input binarization training} 
Starting from the network obtained in stage (a), the inputs are binarized to 0 or 1 before being fed into the network.

\paragraph{Layer-by-layer quantization-aware training}  
In this stage, beginning with the network from stage (b), we sequentially apply quantization-aware training to each layer by constraining weights to the discrete values of \( +1 \), \( -1 \), and 0.

\paragraph{Layer-by-layer gradual pruning and quantization-aware training}  
In this stage, starting from the network of stage (c), we apply gradual pruning while maintaining quantization-aware training on all layers. Ideally, each neuron would retain only the 64 weights with the highest absolute values after each mini-batch, with remaining weights masked to zero in subsequent forward passes. However, directly pruning from 784 to 64 connections, particularly in the first layer, would cause severe underfitting. To mitigate this, we implement a progressive pruning strategy according to Equation \ref{eq:pruning}:
\begin{equation}
C_t = C_i - \frac{(C_i - C_f) \times (s + 1)}{S}
\label{eq:pruning}
\end{equation}
where $C_t$ represents the target connections in the current step, $C_i$ denotes the initial number of connections per neuron in the fully connected state, $C_f$ is the final target number of connections per neuron after pruning, $s$ indicates the current pruning step (starting from 0), and $S$ is the total number of pruning steps. For example, with $S=60$ pruning steps for the first layer, the connections of each neuron gradually decrease from $C_i=784$ to $C_f=64$. This gradual approach preserves critical information while enabling the network to adapt to increased sparsity. Through network pruning, inactive neurons and their corresponding synaptic connections are eliminated, reducing computational cost and enhancing energy efficiency. 

\paragraph{Validation} Finally, the model is evaluated on the entire test dataset.

\subsubsection{Training the Chip Network}
\begin{figure}
    \centering
    \includegraphics[width=0.9\linewidth]{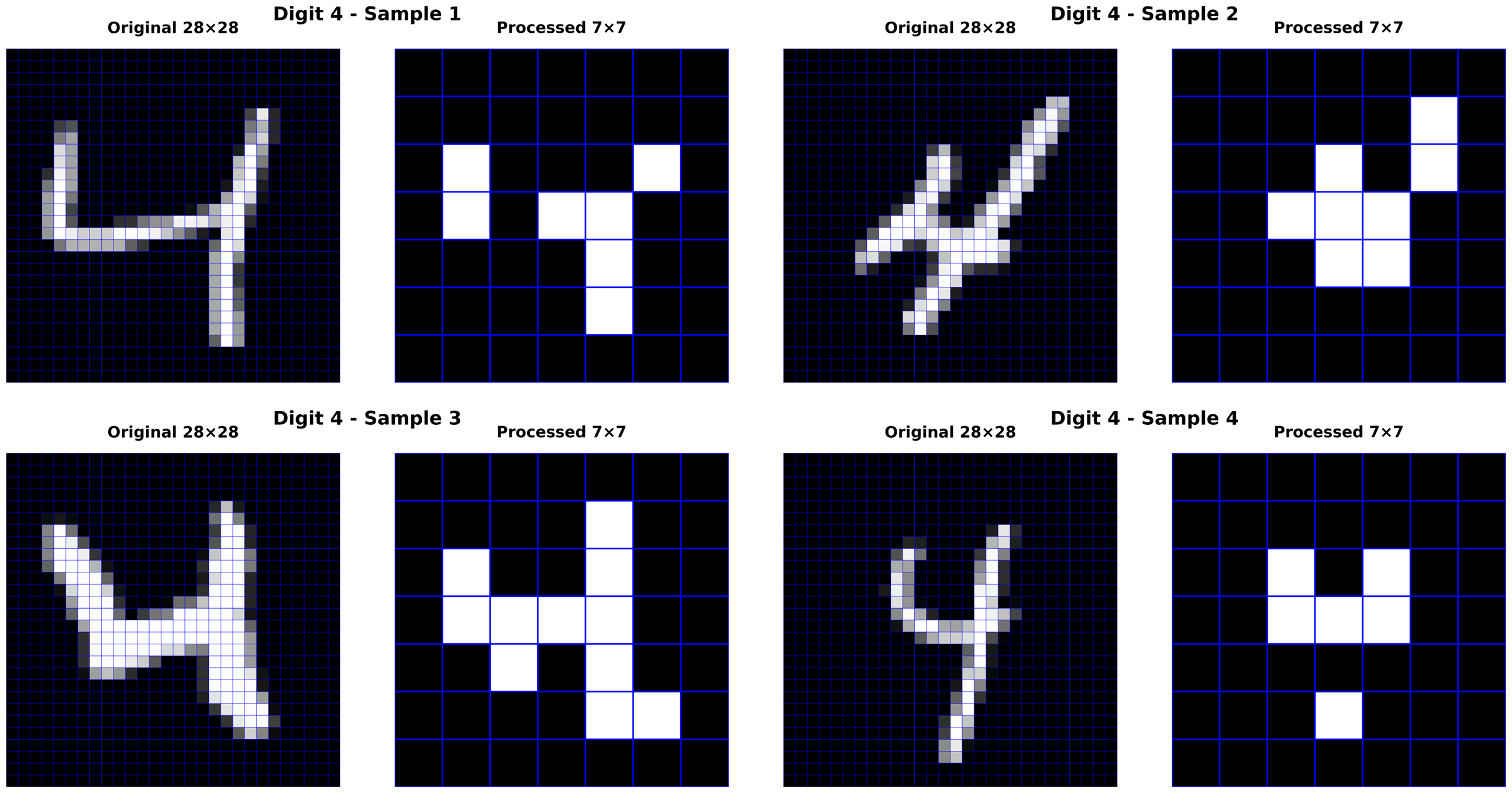}
    \caption{
        Preprocessing for digit 4. Original $28 \times 28$ MNIST images are downsampled to $7 \times 7$ through block averaging with a threshold of 0.3, followed by binarization.
        }
    \label{fig:FigDownsampling4}
\end{figure}

The chip network consists of three layers: an input layer, a hidden layer, and an output layer. The input layer consists of 49 DFFs. The hidden layers contain 24 neurons before pruning, respectively, while the output layer has three neurons. Original \(28\times28\) images are downsampled to \(7\times7\) dimension by block averaging with a threshold of 0.3 and binarization as a preprocessing step, as illustrated in Fig.~\ref{fig:FigDownsampling4}. The weights will still be quantized to \( +1 \), \( -1 \), and 0. This time, our goal is to keep each neuron structurally uniform to utilize standard cells, thereby reducing the burden of layout and routing. After completing the quantization-aware training phase for this mini-network using our previous methodology, we observed that each neuron ultimately retained approximately six positive and two negative weights, a distribution extracted from our highest-accuracy model during training. Based on this observation, we designed each neuron to have six positive and two negative connections, creating a standardized architecture that optimizes computational efficiency and hardware implementation.

During both training and inference, we employ a single forward pass per input sample. At the end of this forward pass, a prediction is deemed correct only if exactly one neuron in the output layer, namely the one corresponding to the correct class, emits a single spike. All other output neurons must remain silent.

The training methodology for the chip network follows a similar approach with four key differences: First, for the loss function, we employ a combination of membrane potential-based loss ($\mathcal{L}_{\mathrm{mem}}$) and spike-based loss ($\mathcal{L}_{\mathrm{spike}}$) with the total loss ($\mathcal{L}_{\mathrm{total}}$) given by \begin{equation} \mathcal{L}_{\mathrm{total}} = 0.6\,\mathcal{L}_{\mathrm{spike}} + 0.4\,\mathcal{L}_{\mathrm{mem}}\, \end{equation} Second, we introduce a weight clamping stage after input binarization training, where weights are restricted to $[-1,1]$. Third, we changed the training sequence by pruning the network first, performing quantization-aware training for the chip network next, and pruning last. Fourth, during the gradual layer-by-layer pruning process, we specifically target a final structure with a maximum of six positive weights and two negative weights per neuron.

\section{Experimental Results}

\subsection{Training Results}

After quantization and pruning, the complete network achieves an accuracy of $96.47\%$ on the complete MNIST dataset. Input data samples are binarized, with each neuron restricted to a maximum fan-in of 64 and weights limited to values of \( +1 \), \( -1 \), and 0. The time step parameter is set to 25. Table~\ref{tab:Full Network Weight Matrix Statistics} presents the weight matrix statistics for the entire network, while Table~\ref{tab:full_network_params} summarizes its architectural parameters.

\begin{table}[htbp]
    \centering
    \caption{Weight Matrix Statistics for the Complete Network}
    \label{tab:Full Network Weight Matrix Statistics}
    \scriptsize
    \setlength{\tabcolsep}{2pt}
    \begin{tabular}{|c|c|c|c|c|}
    \hline
    \textbf{Statistic} & \multicolumn{1}{c|}{\textbf{Hidden}} & \multicolumn{1}{c|}{\textbf{Hidden}} & \multicolumn{1}{c|}{\textbf{Hidden}} & \multicolumn{1}{c|}{\textbf{Output}} \\
    & \multicolumn{1}{c|}{\textbf{Layer 0}} & \multicolumn{1}{c|}{\textbf{Layer 1}} & \multicolumn{1}{c|}{\textbf{Layer 2}} & \multicolumn{1}{c|}{\textbf{Layer}} \\
    \hline
    Size           & 100352           & 12288            & 9216             & 960              \\
    \hline
    Active Neurons  & 128/128   & 96/96     & 96/96     & 10/10     \\ 
    \hline
    1s             & 3957 (3.94\%)    & 2935 (23.89\%)   & 2104 (22.83\%)   & 70 (7.29\%)      \\
    \hline
    -1s            & 3697 (3.68\%)    & 2654 (21.60\%)   & 3410 (37.00\%)   & 273 (28.44\%)    \\
    \hline
    0s             & 92698 (92.37\%)  & 6699 (54.52\%)   & 3702 (40.17\%)   & 617 (64.27\%)    \\
    \hline
    \end{tabular}
    \end{table}

\begin{table}[htbp]
    \centering
    \caption{Complete Network Parameters}
    \label{tab:full_network_params}
    \begin{tabular}{|c|c|}
        \hline
        \textbf{Parameter (Complete Network)} & \textbf{Value} \\
        \hline
        Input pins & 784 \\
        \hline
        Output pins & 10 \\
        \hline
        Layers & 4 \\
        \hline
        Neurons & 330 \\
        \hline
        Max Fanout Hidden Layer0 & 128 \\
        \hline
        Max Fanout Hidden Layer1 & 96 \\
        \hline
        Max Fanout Hidden Layer2 & 96 \\
        \hline
        Max Fanout Hidden Layer3 & 10 \\
        \hline
    \end{tabular}
\end{table}

For the chip network, after quantization and pruning, we achieve the precision $80.07\%$ when classifying the digits $2$, $3$, and $4$. The input data are downsampled from $28 \times 28$ to $7 \times 7$ and binarized, with each neuron having eight fan-in connections (two negative and six positive). The network operates with a single forward pass. Furthermore, we also evaluated various combinations of digits, with digits 0, 1, and 2 producing the highest precision of 86.2\%. These results demonstrate that our Superconductor SNN Chip can effectively process multiple dataset configurations based on hardware configurations.

Finally, we selected digits 2, 3, and 4 for our chip design. The statistics of the weight matrix are presented in Table~\ref{tab:matrix_statistics}, with the values in parentheses representing the percentage of each weight value.
\begin{table}[htbp]
\centering
\caption{Chip Network Weight Matrix Statistics}
\label{tab:matrix_statistics}
\begin{tabular}{|c|c|c|}
\hline
\textbf{Statistic} & \textbf{Hidden Layer} & \textbf{Output Layer} \\
\hline
Total elements & 1176 & 72 \\
\hline
Active Neurons & 18/24 & 3/3 \\
\hline
Number of 1s & 76 (6.46\%) & 18 (25.00\%) \\
\hline
Number of -1s & 21 (1.79\%) & 5 (6.94\%) \\
\hline
Number of 0s & 1079 (91.75\%) & 49 (68.06\%) \\
\hline
\end{tabular}
\end{table}

\subsection{Chip Network Implementation}
\label{sec:finalHardware}
Our superconducting SNN chip has dimensions of 3.4$\times$3.9 mm$^2$ with 40 pins and accommodates a maximum of 25 neurons. Pin allocation includes clock and bias current pins, leaving only seven pins for input data and three pins for output data. Based on hardware constraints, we trained the chip network to fit within the chip's capacity. The overall parameters for the chip network are shown in Table~\ref{tab:chip_params}. 

The hidden layer consists of 18 neurons, reduced from the original 24. This reduction results from two factors: first, the inherent sparsity of the downsampled MNIST dataset, where boundary pixels contain no information (zero values); second, the quantization process, which forces many weight elements to zero. These factors eliminate excitatory connections for certain neurons, rendering them inactive. To accommodate this, we implement a hierarchical structure on our chip. Input data are preprocessed and down-sampled to a $7 \times 7$ format before being fed into the chip. The input layer comprises 49 input shift registers. Due to the limited number of seven input pins, data is shifted in over seven cycles until the entire sample is loaded. Once all 49 shift registers contain the 1-bit data representing a complete sample, individual bias currents are applied to the output fanout circuits, which split and transmit the pulses through PTLs to the fan-in circuits of the hidden layer. After data synchronization, the 18-neuron hidden layer is activated, generating pulses that propagate through their respective fan-out circuits to the fan-in of the output layer (three neurons). The output layer then produces pulses representing the information used for classification. Due to negative clock routing constraints, two cycles are needed to propagate the inputs after loading a complete sample, followed by an additional cycle to retrieve the outputs. This process enables the chip to produce a prediction every ten cycles.


The superconducting SNN chip layout is shown in Fig.~\ref{fig:FigNetwork}.


\begin{figure}[!t]
    \centering
    \includegraphics[width=0.9\linewidth]{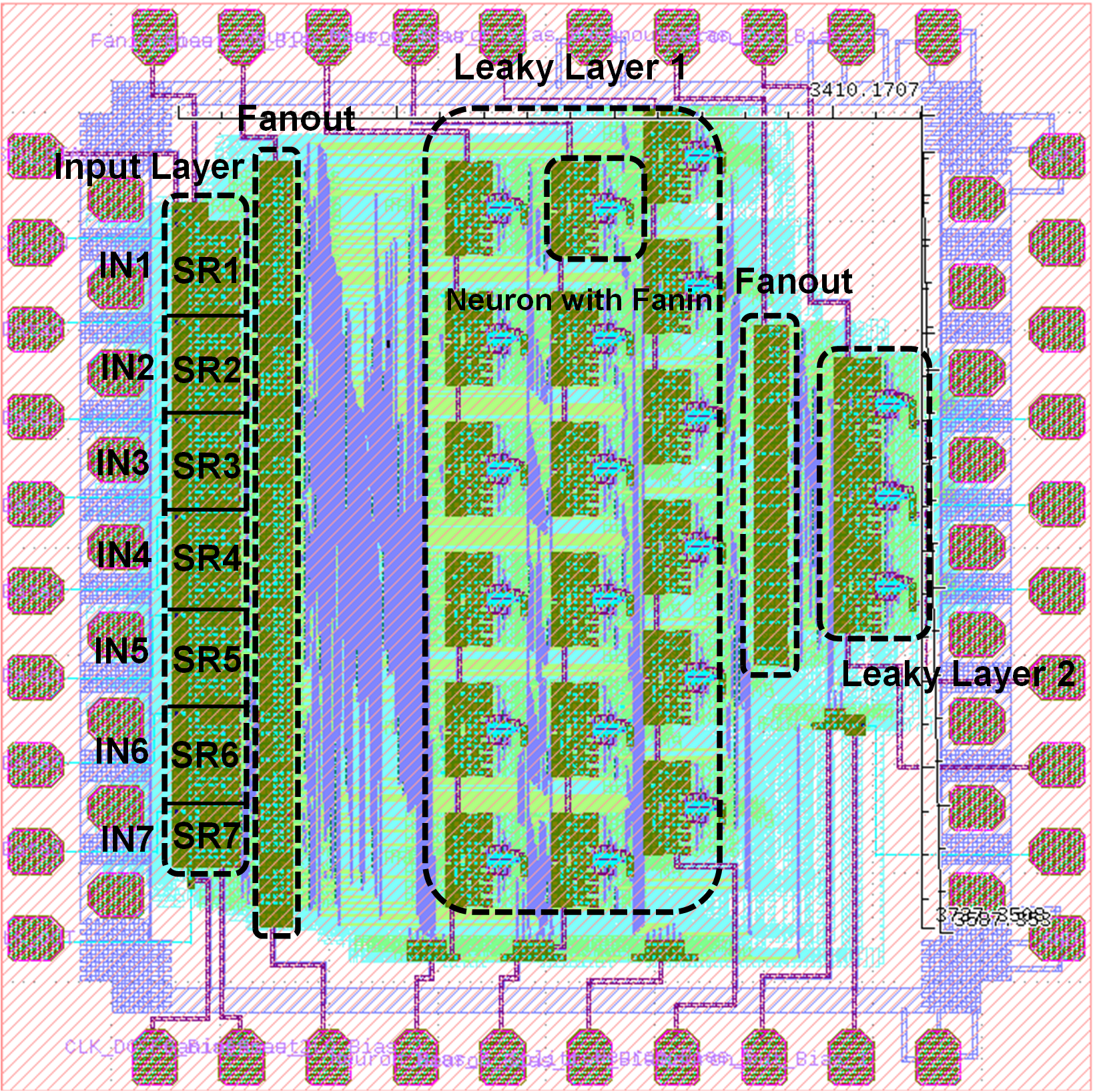}
    \caption{Superconducting SNN Chip}
    \label{fig:FigSNNchip}
\end{figure}

\begin{table}[htbp]
    \centering
    \caption{Chip network parameters for 3 classes of the MNIST dataset.}
    \label{tab:chip_params}
    \begin{tabular}{|c|c|}
        \hline
        \textbf{Parameter (Chip Network)} & \textbf{Value} \\
        \hline
        Input pins & 7 \\
        \hline
        Output pins & 3 \\
        \hline
        Layers & 3 \\
        \hline
        Neurons & 18+3 \\
        \hline
        Max Fanout Hidden Layer & 9 \\
        \hline
        Max Fanout Output Layer & 3 \\
        \hline
    \end{tabular}
\end{table}

\subsection{Power and Accuracy Results}
\begin{figure}
    \centering
    \includegraphics[width=0.8\linewidth]{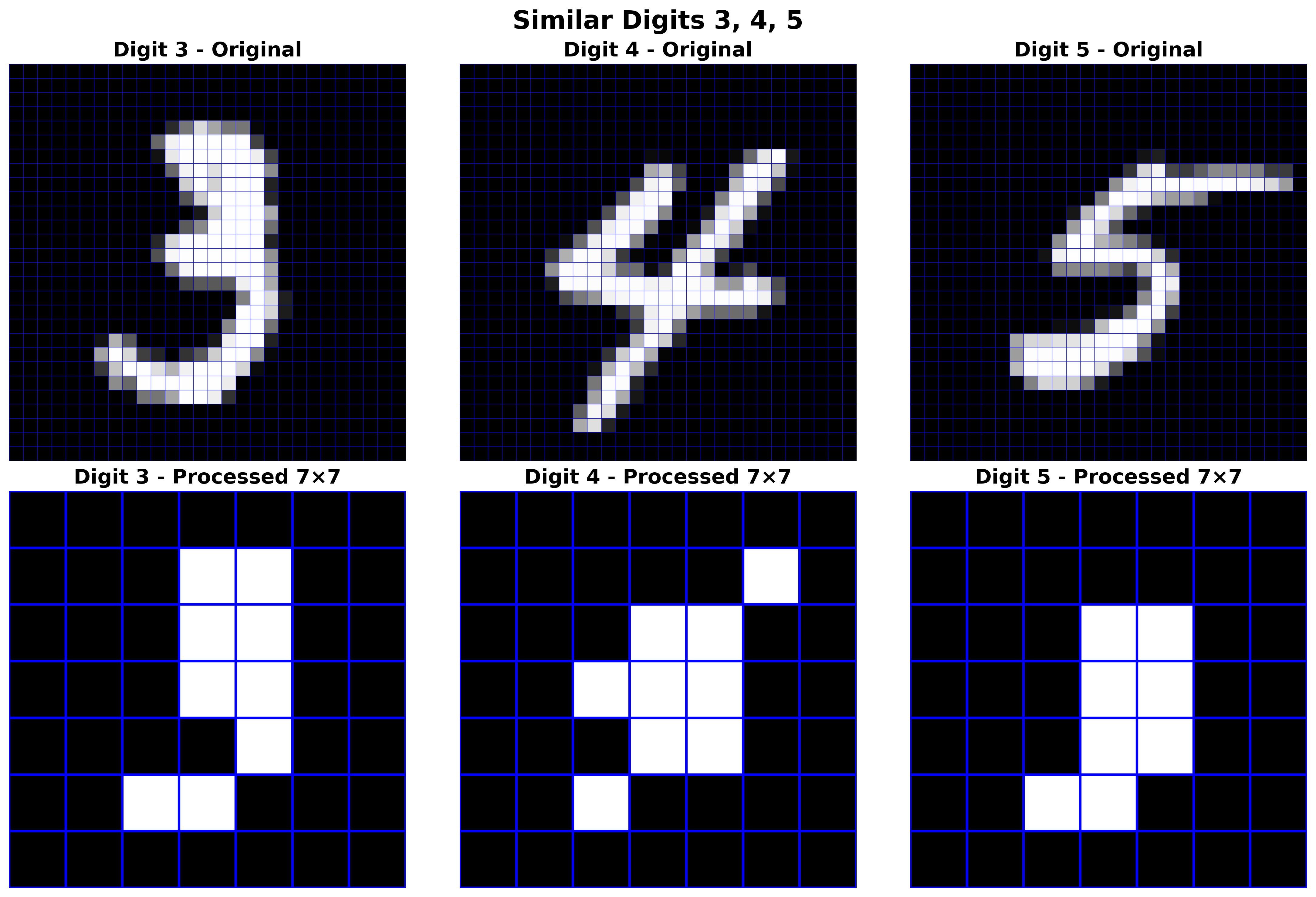}
    \caption{Visual similarity of digits 3, 4, and 5 after $7 \times 7$ downsampling and binarization.}
    \label{fig:Figsimilar_digits_345}
\end{figure}

\begin{figure}
    \centering
    \includegraphics[width=0.8\linewidth]{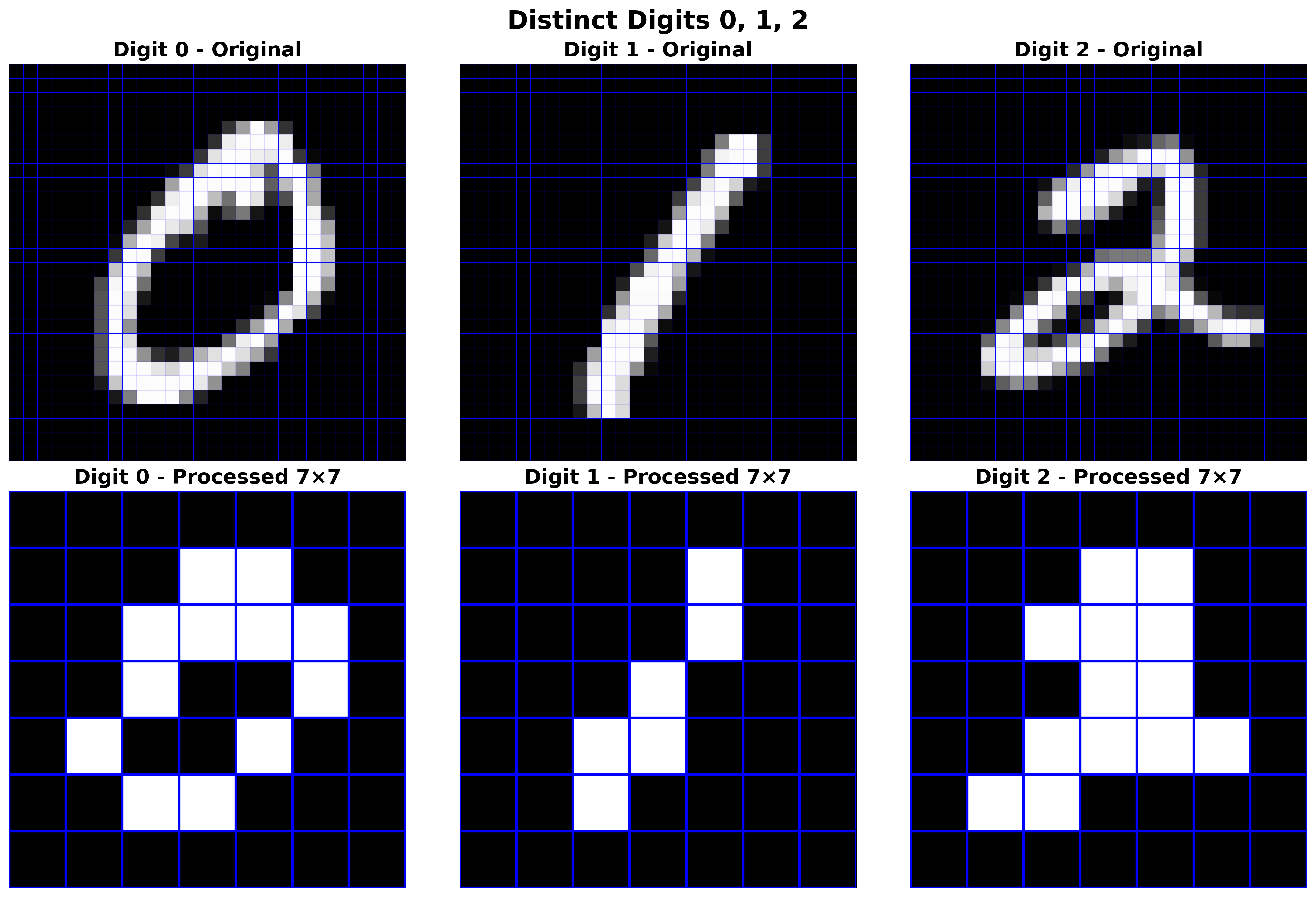}
    \caption{Distinct visual features of digits 0, 1, and 2 after $7 \times 7$ downsampling and binarization.}
    \label{fig:Figdistinct_digits_012}
\end{figure}

We evaluated the performance of our superconducting SNN chip through comprehensive and detailed benchmarking. Table \ref{tab:digit_selection} presents the accuracy results for various digit combinations. Variations in accuracy in different combinations can be attributed to inherent visual similarity between digits after downsampling and binarization. As shown in Figure \ref{fig:Figsimilar_digits_345}, digits '3', '4', and '5' become visually indistinguishable following the $7 \times 7$ downsampling and binarization process, leading to a decrease in the classification accuracy. In contrast, digits '0', '1', and '2' retain more distinctive visual features even after preprocessing, as shown in Figure~\ref{fig:Figdistinct_digits_012}, resulting in higher accuracy. The overall performance could be further enhanced by using higher resolution inputs (e.g. $14 \times 14$) or incorporating convolutional layers for more advanced data preprocessing.

\begin{table}[htbp]
  \centering
  \caption{Resource and power comparison for two of our recent works. The energy for inference is the dynamic energy of JJ switchings.}
  \label{tab:jj-power-comparison}
  \begin{tabular}{|l|c|c|}
    \hline
    \textbf{Metric}               & \textbf{karamuft et al~\cite{Karamuft_Scalable}} & \textbf{This work} \\ \hline
    Predictable digits            & 0--9              & 2, 3, 4 \\ \hline
    Accuracy (\%)                 & 96.1              & 86.2 \\ \hline
    JJs count                     & 196,972           & 5,822 \\ \hline
    Static power (mW)             & 44.6              & 2.15 \\ \hline
    Energy per inference (nJ)     & 1.5               & $1.31\times10^{-6}$ \\ \hline
  \end{tabular}
\end{table}

As summarized in Table \ref{tab:jj-power-comparison}, this work reports a fabricated superconducting SNN prototype that, due to stringent hardware constraints outlined in Section \ref{sec:finalHardware}, focuses on the digit subset 2, 3, and 4, achieving an inference accuracy of 86.2\%. In contrast, the network by Karamuftuglu et al.~\cite{Karamuft_Scalable}, which utilizes high-fan-in neurons and is evaluated in software, classifies all ten MNIST digits with 96.1\% accuracy but requires a significantly larger design (containing 196,972 JJs) and has higher power consumption (44.6 mW static power and 1.5 nJ per inference). Our prototype incorporates 5,822 JJs, consumes 2.15 mW of static power, and has an energy cost of approximately \(1.31\times10^{-6}\) nJ per inference, highlighting the trade-offs involved when transitioning from simulation to a physically realizable superconducting SNN implementation.

\begin{table}[htbp]
    \centering
    \caption{Network accuracy for different digit combinations after downsampling and quantization of the inputs}
    \label{tab:digit_selection}
    \begin{tabular}{|c|c|}
        \hline
        \textbf{Digits} & \textbf{Accuracy (\%)} \\
        \hline
        0, 1, 2 & 86.20 \\
        \hline
        2, 3, 4 & 80.07 \\
        \hline
        3, 4, 5 & 72.34 \\
        \hline
        5, 6, 7 & 75.07 \\
        \hline
    \end{tabular}
\end{table}

\section{Conclusion}
This paper presented a novel design methodology for a superconducting SNN chip based on our innovative superconducting neuron cell. The proposed architecture leverages the event-driven operation of SNNs and the ultra-high-speed, low-power capabilities of SFQ digital circuits to address the inefficiencies inherent in conventional digital systems. We developed a training framework using \texttt{snnTorch}, which incorporates a hybrid loss function that combines membrane potential and spike losses, allowing the network to achieve a classification accuracy of 96.5\% on the entire MNIST dataset after quantification and pruning. To meet the strict hardware constraints of superconducting platforms, such as limited chip area and pin availability, we optimized the network to classify a reduced set of digits, achieving a maximum inference accuracy of 86.2\% for digits 0, 1, and 2 within a chip area of $3.4\times3.9\,\mathrm{mm}^2$. The total JJ count for our chip is 5,822, with a static power consumption of 2.15 mW, and an estimated switching energy of 6.55 fJ per inference. This work demonstrates the potential of superconducting SNNs for high-speed, energy-efficient neuromorphic computing and lays a promising foundation for future developments in scalable neural-hardware systems.



\end{document}